# CONSIDERATIONS ON ENERGY FRONTIER COLLIDERS AFTER LHC*

V. Shiltsev[#], Fermilab, Batavia, IL 60510, USA

*Abstract*

Since 1960's, particle colliders have been in the forefront of particle physics, 29 total have been built and operated, 7 are in operation now. At present the near term US, European and international strategies of the particle physics community are centered on full exploitation of the physics potential of the Large Hadron Collider (LHC) through its high-luminosity upgrade (HL-LHC). The future of the world-wide HEP community critically depends on the feasibility of possible post-LHC colliders. The concept of the feasibility is complex and includes at least three factors: feasibility of energy, feasibility of luminosity and feasibility of cost. Here we overview all current options for post-LHC colliders from such perspective (ILC, CLIC, Muon Collider, plasma colliders, CEPC, FCC, HE-LHC) and discuss major challenges and accelerator R&D required to demonstrate feasibility of an energy frontier accelerator facility following the LHC. We conclude by taking a look into ultimate energy reach accelerators based on plasmas and crystals, and discussion on the perspectives for the far future of the accelerator-based particle physics. This paper largely follows previous study [1] and the presentation given at the ICHEP'2016 conference in Chicago [2].

## INTRODUCTION

Colliding beam facilities which produce high-energy collisions (interactions) between particles of approximately oppositely directed beams have been on the forefront of particle physics for more than half a century and twenty nine reached operational stage [3]. Their energy has been on average increasing by a factor of 10 every decade until about the mid-1990's. Notably, the hadron colliders were 10-20 times more powerful. Since then, following the demands of high energy physics (HEP), the paths of the colliders diverged: to reach record high energies in the particle reaction the Large Hadron Collider was built at CERN, while new $e+e-$ colliders called "particle factories" were focused on detailed exploration of phenomena at much lower energies. The Tevatron, LEP and HERA established the Standard Model of particle physics. The current landscape of the high energy physics is dominated by the LHC. The next generation of colliders is expected to lead the exploration of the smallest dimensions beyond the current Standard Model.

While the development of energy frontier colliders over the past five decades initiated a wide range of innovation in accelerator physics and technology which resulted in 100-fold increase in energy (for both hadron and lepton colliding facilities) and $10^4$-$10^6$ fold increase of the luminosity, the progress in the maximum c.o.m. energy has drastically slowed down since the early 1990's and the lepton colliders even went backwards in energy to study rare processes – see, e.g., Fig.1 in [4]. Moreover, the number of the colliding beam facilities in operation has dropped from 9 two decades ago to 7 now (2016). The future of accelerator-based HEP beyond LHC has been recently debated by several authors in [4-7, 3] and many technical details discussed in the collective book "*Challenges and Goals for Accelerators in the XXI Century*" [8]. Here we bring an economical (financial) perspective to the discussion on feasible colliders beyond the LHC and show that options based on traditional acceleration technologies are very much limited. Only "… technological quantum leaps…will drive the long-term progress of the field. We can expect that these ambitions and far sighted R&D programmes in accelerator technology will redefine the field of high-energy physics in the XXI century…" (M.Mangano, [8], p.21). In general, the discussion on the "beyond the LHC" energy frontier accelerators comes to the question of the right balance between the physics reach of the future facilities and their feasibility which usually assumes the feasibility of their energy reach (whether it is possible to reach the design c.o.m. energy), feasibility of the performance (how challenging is the declared design luminosity) and cost feasibility (is it affordable to build and operate?). While the first two criteria (energy and performance reach) are relatively easy to address on the base of the current state-of-the-art accelerator technology (of, e.g., normal- and superconducting magnets, RF, etc) and beam physics, the feasibility of the cost requires analysis of both the perspective available resources and the facility cost range. In the analysis below we will use the cost of LHC - about 10B$ at today's prices - as a reference for a globally affordable future facility and compare it with the resources required to build "beyond the LHC" colliders, including *"near future"* facilities with possible construction start within a decade - such as the international $e+e-$ linear collider in Japan (ILC) [9] and circular $e+e-$ colliders in China (CepC) [10] and Europe (FCC-ee) [11]; *"future"* colliders with construction start envisioned 10-20 years from now – such as linear $e+e-$ collider at CERN (CLIC) [12], muon collider [13], and circular hadron colliders in China (SppC) [10], Europe (HE-LHC [14] and FCC-pp [11]) and USA (VLHC [15]); and an ultimate *"far future"* colliders with time horizon beyond the next two decades based on beam-plasma [16], laser-plasma [17] and crystal-plasma [18, 4] acceleration technologies.

## COST OF LARGE POST-LHC MACHINES

All large accelerators built so far are based on four major acceleration technologies which employ either normal-conducting RF, or super-conducting RF, or magnets, again normal- or super-conducting. Some used more than one technology at once. All these technologies are well understood and their costs and potentials can be extrapolated from the past experience. In addition, construction of accelerators usually involves civil construction, often – tunnelling, and creating of infrastructure, including high-power electric and cryogenic ones. All these technologies are commercial, in the sense that we just buy corresponding services from industries and, again, their costs are known pretty well. Therefore, one can expect that at least for future accelerators based on these "traditional" accelerator and infrastructure technologies, an extrapolation can be done and a rough cost estimate (better say – cost rage) can be obtained. Indeed, an analysis of the known costs of large accelerator facilities has been undertaken in [19]. Based on publicly available costs for 17 large accelerators of the past, present and those currently in the planning stage it was shown that the "total project cost (TPC)" (sometimes cited as "the US accounting") of a collider can be broken up into three major parts corresponding to "civil construction", "accelerator components", "site power infrastructure". The three respective cost components can be parameterized by just three parameters – the total length of the facility tunnels $L_f$, the center-of-mass or beam energy $E$, and the total required site power $P$ - and over almost 3 orders of magnitude of $L_f$, 4.5 orders of magnitude of $E$ and more than 2 orders of magnitude of $P$ the so-called "$\alpha\beta\gamma$-cost model" works with ~30% accuracy [19]:

$$TPC \approx \alpha \times (Length/10km)^{1/2} + \beta \times (Energy/TeV)^{1/2} + \gamma \times (Power/100MW)^{1/2} , \quad (1)$$

where coefficients $\alpha=2B\$/(10km)^{1/2}$, $\gamma=2B\$/(100MW)^{1/2}$, and accelerator technology dependent coefficient $\beta$ is equal to $10\ B\$/TeV^{1/2}$ for superconducting RF accelerators, $8\ B\$/TeV^{1/2}$ for normal-conducting ("warm") RF, $1B\$/TeV^{1/2}$ for normal-conducting magnets and $2B\$/TeV^{1/2}$ for SC magnets (all numbers in 2014 US dollars).

Let's take the LHC as an example. The first component of the "$\alpha\beta\gamma$-model" is the cost of some 40 km of LHC tunnels (including 27 km of the LEP tunnel, 7 km of SPS, injectors and beamlines) which can be estimated as $2B\$\times(40/10\ km)^{1/2}=4B\$$. The estimate of the second component is dominated by the cost of SC magnets for 14 TeV com collider, i.e., $2B\$\times(14)^{1/2}=7.5B\$$. Finally, the estimate of the 150 MW power infrastructure piece is $2B\$\times(150\ MW/100MW)^{1/2}=2.5B\$$, that makes the TPC range of the LHC – if built from scratch - equal to $4B\$+7.5B\$+2.5B\$= 14B\$ \pm 4.5B\$$. The CERN LHC Factbook [20] indicates the cost of LHC project of 6.5BCHF, including 5BCHF for accelerator facility.

These numbers are in so-called "European accounting"- the different methodology of the cost estimates widely used in Europe – that includes only the industrial contracts for major items like civil engineering, the accelerator elements and corresponding labor requirements (such approach is often referred). Usually, the "European accounting" is factor of 2.0–2.5 lower than the US DOE Office of Science's "the total project cost" (TPC) accounting [19] which additionally includes the costs of the required R&D, development of the engineering design, project management, escalation, contingency, overhead funds, project-specific facility site development, sometimes - detectors, etc. Therefore, the TPC of the LHC accelerator project is some $5BCHF\times(2-2.5)=10-12.5\ BCHF=10-12.5\ B\$$. Add an estimated 3-4B\$ for the LHC injector complex needed if the LHC would be built as a "green field" and one gets the LHC TPC of about $13-16.5B\$$ - very much in line with the "$\alpha\beta\gamma$--estimate" we obtained above.

Similar kind of estimates have been done many facilities currently considered for post-LHC HEP accelerators – see, e.g, [21] – and summarized in Fig.1 below.

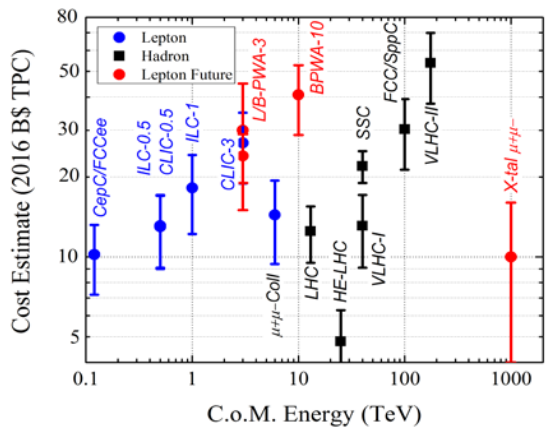

Figure 1: Center-of-mass beam energy vs estimated Total Project Cost (TPC, the "US Accounting", in B$) of various post-LHC frontier HEP accelerators: (blue dots) – lepton colliders, (black dots) – hadron colliders, (red dots) – future linear lepton colliders (see text).

One can see that only HE-LHC is certainly financially feasible being below the 10B$ "feasibility level" discussed above. Several machines are potentially within the financial reach if undertaken as a global HEP project  - CepC, FCC-ee, 0.5 TeV options of ILC and CLIC, 6 TeV c.o.m. muon collider and possibly, low-field option of the large circumference (233km) VLHC-I. At the same time, really questionable seems to be the 1 TeV version of the ILC, 3 TeV CLIC, 87-km long SSC and 60-100 km FCC pp and SppC.

## DISCUSSION AND SUMMARY

Several ways to assure feasibility of future colliders are being considered. Firstly, significant savings can be achieved by re-using the existing infrastructure and/or existing accelerators as injectors for the future ones. Of course, that can not be applied universally (eg., not for frontier lepton colliders), but, for example, CERN's proton accelerator complex can be used in the FCC-pp or a muon collider. Secondly, launching extensive R&D programs focused on the cost reduction of traditional technologies, e.g., SC magnets and tunnelling, can greatly help, too [22]. At the same time, one should take into account that this approach has its limits. Thirdly, one could try to benefit significantly lower cost of doing business in Asia, particularly, in China – for example, comparison of modern synchrotron light sources shows a factor of about 3 lower construction cost for comparable facilities [22]. This advantage may or may not be in effect in the future but it definitely should be taken into account.

While discussing feasibility of future machines, one should also take into consideration expected long period of commissioning and the issue of availability of accelerator experts needed for construction and operation of large(r) colliders. Indeed, due to high complexity of modern colliders, attainment of their design or the ultimate luminosity can take quite substantial time [23]. The latest example is LHC, where it took almost 8 years to get from the first circulating beams (September 10, 2008) to the design luminosity of $10^{34}$ cm$^{-2}$s$^{-1}$ (July 6, 2016). The issue of availability of experts can be illustrated by the example of the ILC project which estimates that some 13,000 man-years (FTEs, or full-time-equivalent) of accelerator scientists, engineers and technicians are needed over some 8 years of construction of the International Linear Collider [9] – that gives on average 1,600 trained people needed for installation, integration, testing and quality assurance, commissioning and all other related activities associated with 7.8B$ worth of materials and services budget. Despite the lack of a crisp definition of who should be considered "an accelerator expert", one can estimate that the world-wide community of accelerator physicists and experienced engineers does not exceed 1200-1500 people and the total accelerator personnel (all scientists, engineers, technicians, drafters, etc) is about 4,000-4,500. Therefore, any plans for a really big facility at the scale of 10B$ should take into account that significant time will be needed to get the required number of the people together.

The most promising option in the long run might be to develop a new accelerator technology, namely, ultra-fast plasma wake-field acceleration (PWFA). The potential of the method is enormous, though recent attempts to design a collider based on laser- or beam-driven PWFA [16, 17] showed many serious not yet resolved issues such as modest average accelerating gradient (~2 GeV/m vs maximum single stage value of 10-50 GeV/m), uncertain effectiveness of staging, low luminosity as the result of the beam emittance growth due to scattering of electrons and positrons in plasma, extremely tight tolerances on transverse and longitudinal stability of the collider elements, (currently) very low efficiency of the electric plug power conversion to beam power, etc. Cost-wise, such colliders are not very efficient at the present stage of development – see red dots in Fig. 1 and discussion in [19] – but, again, they have a significant potential for cost savings, for example, due to quick reduction of the cost per Watt of pulsed high power lasers.

One can try to look into options for "ultimate" future energy frontier collider facility with c.o.m. energies of 300-1000 TeV (20-100 times the LHC). We surely know that for the same reason the circular $e+e-$ collider energies do not extend beyond the Higgs factory range (~0.25 TeV), there will be no circular proton-proton colliders beyond 100 TeV because of unacceptable synchrotron radiation power – they will have to be linear. It is also appreciated that even in the linear accelerators electrons and positrons become impractical above about 3 TeV due to beam-strahlung (radiation due to interaction at the IPs) and about 10 TeV due to radiation in the focusing channel (<10 TeV). This leaves only $\mu+\mu-$ or $pp$ options for the "far future" colliders. If we further limit ourselves to affordable options and request such a flagship machine not to exceed $L_f$ ~10 km in length then we seek a new accelerator technology providing average gradient of >30 GeV/m (compare with $E/L_f$~ 0.5 GeV per meter in the LHC). There is only one such option known now: super-dense plasma as in, e.g., crystals [18], that excludes protons because of nuclear interactions and leaves us with muons as the particles of choice [4]. High luminosity can not be expected for such a facility if we limit the beam power and, with necessity, the total facility site power to some affordable level of $P$ ~100MW. Indeed, as the energy of the particles $E$ grows, the beam current will have to go down at fixed power $I=P/E$, and, consequently, the luminosity will need to go down with energy. The paradigm shift from the past collider experience when luminosity scaled as $L \sim E^2$ will need to happen in the "far future" of HEP.

To summarize, a short answer to the question *"Will There Be Energy Frontier Colliders After LHC?"* from the accelerator stand point is "may be". Longer answer would include a notion that such a collider will first need a strong motivation for conctruction, i.e. be dependent on the LHC results. If based on "traditional" accelerator technologies (SRF, SCMag, etc), only HE-LHC is cost feasible (will cost about half of that of the LHC), few others are close to the LHC TPC or exceed it slightly - CepC/FCCee, ILC, Muon Coll, VLHC-I, other proposed facilities would need either significant R&D or/and use advantages of developing economy in China. The hopeful "non-traditional" technology of plasma acceleration is very expensive now, it needs several decades of R&D to prove its feasibility as viable post-LHC collider option. On the other hand it has great potential and can be a basis of an "ultimate (dream)" $O$(1PeV) crystal muon collider which by necessity will be low luminosity and will require a paradigm change for HEP reaserch.